\documentclass[twocolumn,aps,pre,amsmath]{revtex4}
\usepackage{graphicx}

\newcommand{\s}{\sigma}
\newcommand{\tw}{t_{\rm w}}
\newcommand{\Tc}{T_{\rm c}}
\newcommand{\Teff}{T_{\rm eff}}
\newcommand{\cnl}{C_{\rm nl}}
\newcommand{\rnl}{R_{\rm nl}}

\newcommand{\cg}{C_{\rm g}}
\newcommand{\rg}{R_{\rm g}}
\newcommand{\xg}{X_{\rm g}}
\newcommand{\cl}{C_{{\rm loc}}}
\newcommand{\rl}{R_{{\rm loc}}}
\newcommand{\xl}{X_{{\rm loc}}}
\newcommand{\wcl}{\widetilde{C}_{{\rm loc}}}
\newcommand{\wrl}{\widetilde{R}_{{\rm loc}}}
\newcommand{\wxl}{\widetilde{X}_{{\rm loc}}}
\newcommand{\cln}{C_{{\rm loc,0}}}
\newcommand{\rln}{R_{{\rm loc,0}}}
\newcommand{\xln}{X_{{\rm loc}}}
\newcommand{\clo}{C_{{\rm loc,1}}}
\newcommand{\rlo}{R_{{\rm loc,1}}}
\renewcommand{\th}{\mbox{th}}
\newcommand{\half}{\frac{1}{2}}
\newcommand{\eql}{_{\rm eq}}
\newcommand{\hext}{h^{\rm ext}}
\renewcommand{\d}{b}

\newcommand{\lav}{\langle}
\newcommand{\rav}{\rangle}
\newcommand{\be}{\begin{equation}}
\newcommand{\ee}{\end{equation}}
\newcommand{\bea}{\begin{eqnarray}}
\newcommand{\eea}{\end{eqnarray}}
\newcommand{\nn}{\nonumber}
\newcommand{\eq}[1]{~(\ref{#1})}
\newcommand{\qv}{{\bf q}}
\newcommand{\rv}{{\bf r}}
\newcommand{\Cq}{C_\qv}
\newcommand{\Rq}{R_\qv}
\newcommand{\omq}{\omega_\qv}
\newcommand{\aq}{a_\qv}
\renewcommand{\rq}{r_\qv}
\newcommand{\dq}{\int(dq)\,}
\newcommand{\zz}{\d}
\newcommand{\scc}{{\mathcal F}_{C}}
\begin{document}

\title{Universality of Fluctuation-Dissipation Ratios: The Ferromagnetic Model}
\author{A.Garriga$^1$, P. Sollich$^2$, I. Pagonabarraga$^3$ and
F. Ritort$^3$}
\affiliation{$^1$ Departament de Tecnologia, Universitat Pompeu
Fabra.  Passeig de Circumval.laci\'o 8, 08003 Barcelona, Spain\\ $^2$
King's College London, Department of Mathematics, Strand, London WC2R
2LS, U.K.\\ $^3$ Departament de Fisica Fonamental, Facultat de
F\'{\i}sica, Universitat de Barcelona, Diagonal 647,
08028 Barcelona, Spain}
%E-Mail: adan.garriga@upf.edu, ritort@ffn.ub.es,ipagonaabrraga@ub.edu}
\date{\today}

\begin{abstract}
We calculate analytically the fluctuation-dissipation ratio (FDR) for
Ising ferromagnets quenched to criticality, both
for the long-range model and its short-range analogue in the limit of large
dimension. Our exact solution
shows that, for both models,
$X^\infty=1/2$ if the system is unmagnetized while
$X^\infty=4/5$ if the initial magnetization is non-zero.
This indicates that two different
classes of critical coarsening dynamics need to be distinguished
depending on the initial conditions, each with its own nontrivial
FDR. We also analyze the dependence of the FDR on whether local
and global observables are used. These results clarify how a proper
local FDR
(and the corresponding effective temperature) should be defined in
long-range models in order to avoid spurious inconsistencies and maintain the
expected correspondence between local and global results; global
observables turn out to be far more robust tools for detecting
non-equilibrium FDRs.
\end{abstract}
\maketitle

\section{Introduction}
\label{sec:intro}

One of the main goals of modern statistical mechanics is to find a
general theory of non-equilibrium processes. Although significant
advances have been made in the past~\cite{mazur,kubo}, a complete
theory of non-equilibrium systems analogous to thermodynamics does not
yet exist.  A
common approach has been to extend well-known equilibrium concepts to
the non-equilibrium regime. One of the most important among these is the
temperature, and many researchers have therefore tried to extend it to systems
out of equilibrium by introducing a so-called \textit{effective
temperature}~\cite{jou}. The question of whether and how such a
quantity can be 
defined properly 
is a central issue in the construction of a general
theory of non-equilibrium systems. Different definitions have been employed in
granular systems~\cite{edwards}, driven systems~\cite{kurreo} and
glassy systems~\cite{kurchan_peliti} among others. Studies of
mean-field spin-glasses have shown that an effective temperature
can be defined by measuring the violation of the
fluctuation-dissipation theorem (FDT) in terms of
a so-called fluctuation-dissipation ratio
(FDR)~\cite{kur_sg}
%
%Glassy systems are systems which relax so slowly~\cite{sitges}
%that effectively live out-of-equilibrium. During very large time
%intervals, some measurable quantities, such as the energy, remain
%practically constant. However, if we analyze the relaxation
%processes (studying the magnetic susceptibility for example) we
%can see that glassy materials are non-stationary and display aging
%\cite{bou1}. The slow relaxation of glassy materials make them the
%perfect laboratory to test new NE theories.
%
\be X(t,\tw)=\frac{TR(t,\tw)}{\frac{\partial
C(t,\tw)}{\partial \tw}}
%\label{eq:fdr_intro}
\nn
\ee
where $T$ is
the temperature of the heat bath,
$C(t,\tw)$ is the autocorrelation function of a given
observable and $R(t,\tw)$ the conjugate response function; the latter
encodes the change of the value of the observable at time $t$ to a
small perturbation at an earlier time $\tw$. 
In equilibrium FDT is verified and
$X(t,\tw)=1$. Whether the FDR is useful more generally, and in
particular beyond mean-field models, has been the subject
of debate in recent years~\cite{teo,soll1}. It has been shown
that the effective temperature defined through the FDR, $\Teff
(t,\tw)= T /X(t,\tw)$, has good thermodynamic properties for some
mean-field models~\cite{kurchan_peliti}; the zeroth law
of thermodynamics can also be extended to the non-equilibrium
regime~\cite{art1}. Nevertheless, there are still many open questions 
regarding the physical meaning and universality of the
FDR~\cite{felix_cris,cugli}. 

An important aspect in the study of FDT violation in glassy systems is
its application to ferromagnetic systems which are quenched from high
temperature to the critical temperature (see
e.g.\ Refs.~\cite{luck1b,soll2,calabrese} and the recent
review~\cite{CalGamreview}) or 
below. The ensuing non-equilibrium evolution, where the
system coarsens -- by the growth of domains with the equilibrium
magnetization, for $T<T_c$ -- is of course different from that of
glasses in many
respects; for example, thermal activation effects are irrelevant for
the long-time dynamics. However, there are also appealing
similarities. In particular, equilibrium is never reached in an
infinite system and as a consequence the system exhibits aging, i.e.\
a dependence of the relaxation properties on the time elapsed since
the quench.

The simplest ferromagnetic model that can be studied at criticality is the
Ising chain first solved by Glauber~\cite{glauber}. At the critical
point, $T=0$, the magnetization jumps discontinuously from 0 ($T>0$)
to 1. The relaxation
dynamics at zero temperature after a quench from $T=\infty$, i.e.\ a
random initial configuration, has been studied in
e.g.\ Refs.~\cite{bray,rieger,prados}. Recently, the FDR has been
calculated analytically
for a randomly staggered perturbation and the
corresponding spin autocorrelation function~\cite{luck1a,lipiello}. For long times $t$ and
$\tw$ this gives $X(t,\tw)=(1+{\tw}/t)/2$. In the limit
$t \to \infty$ the FDR then approaches $X=1/2$,
which coincides with the value 
obtained in models characterized by diffusive dynamics (such as the
random walk or the Gaussian model~\cite{parisi}). These results have
lead to the suggestion~\cite{luck1b,luck2} that for systems at criticality
the limiting value of the FDR
\be
X^\infty=\lim_{\tw\to\infty}\lim_{t\to\infty}X(t,\tw)
%\label{eq:fdr}
\nn\ee
is a universal quantity. Consistent with this, the exact solution of
the ferromagnetic spherical model in $d$ dimensions at criticality
also gives $X^\infty=1/2$ for $d>4$, i.e.\ above the upper critical
dimension~\cite{luck1b,luck2}. 

In considering the universality of the limiting FDR, one issue is
whether and how the limiting FDR depends on the observable whose
correlation and response are measured. An obvious alternative to the
local spin autocorrelation is its long-wavelength analogue, i.e.\ the
correlation function of the fluctuating magnetization. Exact
calculations for the Ising chain~\cite{MaySol,soll2} and the spherical
model~\cite{spherical_model_forthcoming} as well as numerical
simulations~\cite{soll2,comm} for 
the Ising model in $d=2$ show that the resulting
global $X^\infty$ is always identical to the local version. This
local--global correspondence, which can also be obtained by
field-theoretic arguments~\cite{calabrese,CalGamreview}, is rather
reassuring; physically, it
arises because the long wavelength Fourier components of the spins are
slowest to relax and dominate the long-time behaviour of both local
and global quantities. 
For a numerical determination of the limiting FDR the global
quantities are often more suitable~\cite{soll2,comm} because plots of
susceptibility (integrated response) versus correlation are close to
straight lines with slope $X^\infty$. We do not discuss in this paper the FDR
for other observables that are nonlinear in the spins, e.g.\ the
energy; this question is studied
in~\cite{calabrese,CalGamreview,spherical_model_forthcoming}.

A further key question is to what extent the limiting FDR $X^\infty$
depends on the initial conditions for the coarsening dynamics. The
results quoted above all apply to initial high-temperature
equilibrium, i.e.\ an unmagnetized system with no or only short-range
spatial correlations. Some initial progress in considering more
general initial states has already been made. For the Ising
chain~\cite{PicHen02} a non-zero initial magnetization does not change
the value of $X^\infty$; this is unlikely to be a general result,
however, because the Ising chain at its $T=0$ critical point has the
peculiarity that the magnetization remains constant instead of
decaying to zero. Other values of the limiting FDR are nevertheless
possible even in the Ising chain; e.g.\ when the magnetization is zero
initially but correlations between spins are so strong that only a
finite number of domain walls exist in the system, one
finds~\cite{henkel} $X^\infty=0$. A more general analysis for
unmagnetized but spatially correlated initial conditions in the
spherical model gives qualitatively similar results~\cite{PicHen02}:
if initial correlations are strong (i.e.\ decay with distance as slow
power laws), the limiting FDR is $X^\infty=0$; otherwise it is the
same as for an uncorrelated initial state.

In this paper we investigate in full detail the dependence of the FDR
on the initial condition for a ferromagnet with long-range
interactions as well as short-range interactions in the limit of large
dimension $d$. Our main result will be that the non-equilibrium
dynamics starting from unmagnetized and magnetized initial conditions
are in different universality classes that are distinguished by
different and nontrivial, i.e.\ non-zero, values of $X^\infty$.  We
also analyze the correspondence between local and global FDRs. In a
naive analysis this appears broken in the long-range case, but we show
that it is recovered when finite-size corrections are included.

The paper is organized as follows. Section \ref{sec:infinite_N}
describes in outline the calculation of the basic evolution equations
for correlation and response in the long-range ferromagnet. In section
\ref{sec:fdr} we compute from these the global and local FDRs.
Section \ref{sec:finite} contains the corresponding analysis for the
short-range ferromagnet, where the lengthscale-dependence of the FDR
can be made explicit and the local--global correspondence holds as
expected. In section \ref{sec:corrections} this correspondence is
shown to hold also for the long-range model once finite-size
corrections are accounted for. Section \ref{sec:summary}, finally,
summarizes our key results and conclusions. Technical details are
relegated to two appendices.

\section{Long-range ferromagnet}
\label{sec:infinite_N}

In this section we study the ferromagnet with long-range
interactions and in particular the
non-equilibrium dynamics of the relaxation functions (correlations
and responses). We leave technical details to Appendix
\ref{app:technical} and focus here on the conceptual aspects of the calculation
and the results.

The model is defined by the Hamiltonian
\be H=-\frac{J_0}{N-1}\sum_{i,j} \s_i\s_j - \sum_i \hext_i\s_i
\label{eq:ham}\ee
where the $\s_i=\pm 1$, $i=1\ldots N$, denote Ising spin variables
and $\hext_i$ is a position-dependent external field.
The strength of the coupling
between spins, $J_0$, is normalized by a factor $1/(N-1)$ to
ensure an extensive energy; we will take
$J_0=1$ without loss of generality. The dynamics we consider is of
Glauber type: each spin $\s_i$ flips independently with rate
$[1-\s_i \tanh(\beta h_i)]/2$ where $\beta=1/T$ is the inverse
temperature and $h_i$ the local field
acting on spin $i$,
\be h_i = \hext_i + \frac{1}{N-1}\sum_{j\neq i}\s_j.
\label{eq:field}\ee
To keep the notation compact we define the function
$\th(z)=\tanh(\beta z)$ and abbreviate $t_i=\tanh(\beta
h_i)=\th(h_i)$. Multiplying the flip rate by the change $-2\s_i$
when $\s_i$ flips, it follows that
\be
\frac{\partial }{\partial t}\lav \s_i\rav = \lav t_i-\s_i\rav
\label{eq:si_tt}
\ee
where the brackets denote an average over the thermal history of the
system and over the initial conditions. For spin products
($i\neq j$) one finds similarly
\be
\frac{\partial }{\partial t}\lav \s_i \s_j \rav = \lav (t_i-\s_i)\s_j\rav
+ \lav \s_i(t_j-\s_j)\rav.
\label{eq:sij_tt}
\ee
Throughout this paper, omitted time arguments indicate
dynamical averages evaluated at time $t$. To obtain the two-time
correlation functions we can take advantage of the fact  that\eq{eq:si_tt} does
not depend on the initial condition and so is equally valid for correlations
with some quantity at an earlier time $\tw<t$. This gives
\be \frac{\partial }{\partial t}\lav \s_i(t) \s_j(\tw)\rav =
\lav[t_i(t)-\s_i(t)]\s_j(\tw)\rav.
\label{eq:sij_ttw}
\ee
Using \eq{eq:si_tt} we can also express the time evolution of the
magnetization, $m=N^{-1}\sum_i 
\lav \s_i\rav $, as
\be
\frac{\partial m}{\partial t} = -m + \frac{1}{N}\sum_i \lav t_i\rav.
\label{eq:m_t}
\ee
%
%Assuming spatial-translation invariance, which in the long-range
%case considered here means that all spins are equivalent, one also
%has $m=\lav \s_j\rav$ for any $j$ and correspondingly the second
%term in\eq{eq:m_t} can be replaced by $\lav t_j\rav$.
%
%We consider the connected correlation functions defined by,
%
%\be C_{ij}(t,\tw) = \lav \s_i(t)\s_j(\tw)\rav - \lav \s_i(t)\rav
%\lav \s_j(\tw)\rav~~~~. \ee
%
%First we analyze the equal time correlation functions. At equal
%times, the local correlations are trivially,
%
%\be C_{ii}(t,t) = 1 - m^2(t)~~~~, \label{eq:Cii_tt} \ee
%
The relaxation dynamics of physical systems is routinely characterized
by correlation and response functions.  We
will consider here the properties of both global and local relaxation
functions. Our main goal is to clarify the relationship between them
and the implications for the corresponding FDRs. Note that while
the fluctuations of individual spins, as encoded in the local
\mbox{(auto-)} correlation function, are $O(1)$, those of the
fluctuating magnetization $N^{-1}\sum_i \s_i $ are
$O(N^{-1/2})$. Nevertheless we will see that the same physics can be
extracted from both quantities. The global correlation function is
scaled by a factor of $N$ below to give an $O(1)$ quantity as in the
local case.
%
% The main
%difference among them is that while local correlations and responses
%are of order
%$O(1)$,  global ones are of order
%$O(N^{-1})$. We will see how to obtain the same physics from both
%approaches.

Previous studies~\cite{felixsoll,SolFieMay02} have shown that the relevant correlation functions are the connected ones, defined by
\be C_{ij}(t,\tw) = \lav \s_i(t)\s_j(\tw)\rav - \lav \s_i(t)\rav
\lav \s_j(\tw)\rav. \nn\ee
First we analyze the equal-time correlations. For $i\neq j$ it
follows from\eq{eq:sij_tt} that
\bea
\frac{\partial }{\partial t}C_{ij}(t,t) &=&
\lav(t_i-\s_i)\s_j\rav + \lav\s_i(t_j-\s_j)\rav
\nonumber\\
& &
{}-{}\lav t_i-\s_i\rav \lav\s_j\rav - \lav\s_i\rav \lav t_j-\s_j\rav
\nn\\
&=& -2C_{ij}(t,t) + \lav \Delta t_i \Delta \s_j \rav + \lav \Delta
\s_i \Delta t_j \rav ~~~~. \label{eq:Cij_tt} \eea
where we use the notation $\Delta \psi = \psi - \lav \psi\rav$ for the
deviation of any quantity from its average, so that e.g.\ 
$\Delta t_i=t_i-\lav t_i\rav$ and similarly for $\Delta\s_i$. For the
two-time correlations one gets similarly from\eq{eq:sij_ttw}:
\bea
\frac{\partial }{\partial t} C_{ij}(t,\tw) &=&
\lav[t_i(t)-\s_i(t)]\s_j(\tw)\rav 
\nonumber\\
& &{}-{} \lav[t_i(t)-\s_i(t)]\rav \lav\s_j(\tw)\rav
\nn
\\
&=& - C_{ij}(t,\tw) + \lav \Delta t_i(t)\Delta \s_j(\tw)\rav.
\label{eq:Cij_ttw}
\eea
Equations\eq{eq:Cij_tt}
and\eq{eq:Cij_ttw} are general and also valid for short-range systems
provided that the appropriate local field replaces the long-range
expression\eq{eq:field}.

To make progress, we
exploit the fact that in the long-range model, for large $N$, the
correlations between different spins are of $O(N^{-1})$. So the
fluctuations $\Delta h_i$ of $h_i$ around its mean $\hext_i+m$ are
small, of $O(N^{-1/2})$. In equilibrium away from criticality one
can indeed show that $\lav (\Delta h_i)^2 \rav = O(N^{-1})$ while
$\lav (\Delta h_i)^2\Delta \s_j\rav$ and $\lav (\Delta h_i)^3\rav$
are $O(N^{-2})$. We will assume that the correlations out of
equilibrium are of the same order. This is reasonable if we start
from an initial state with weak correlations and, for quenches to
criticality, also restrict
ourselves to the interesting non-equilibrium regime where all
times are short compared to the equilibration time.

Setting the external field to zero, we can Taylor expand the nonlinear
terms in our equations of motion in powers of $\Delta h_i$:
\be
t_i = \th(m) + \Delta h_i\th'(m)
 + \half (\Delta h_i)^2 \th''(m)  + \ldots
\label{t_i_expansion}
\ee
%
%and so
%%
%\be
%\Delta t_i = \Delta h_i \th'(m) + [(\Delta h_i^2)
%-\half \lav(\Delta h_i)^2\rav] \th''(m) + \ldots
%\ee
%%
Since we are only interested in the leading terms, we truncate this
expansion after the linear term in $\Delta h_i$; subleading
corrections are discussed in Sec.~\ref{sec:corrections}. Since $\lav
\Delta h_i \rav = 0$, the leading order term in the equation of motion
for the magnetization\eq{eq:m_t} takes the expected mean-field
form
\be \frac{\partial m}{\partial t} = -m + \th(m).
\label{eq:m0_t} \ee
Due to spatial translation invariance (or more precisely permutation
invariance) between spins in the
long-range ferromagnet, there are only two different correlation
functions, one local and one non-local. We write these as
\be C_{ii} = \cl + O(N^{-1})\label{eq:local} \ee
\be C_{ij} = \cnl/N + O(N^{-2})\label{eq:nonlocal}\ee
using the fact that the non-local correlations are only
$O(1/N)$ to leading order.

In terms of the local\eq{eq:local} and non-local\eq{eq:nonlocal}
correlations, the global correlation is defined by
\be \cg\equiv\cl+\cnl\label{eq:cg} \ee
and gives the leading contribution of the correlator of the
magnetization; see eq.\eq{eq:Cij} of Appendix \ref{app:technical}.
In order to compute the dynamical equation for the global
correlation function\eq{eq:cg} we need the equations for the local
and non-local correlation functions, which can be expressed as
\be \frac{\partial}{\partial t}\cnl(t,t) = -2a\cnl(t,t) +
2\th'(m)(1-m^2) \label{eq:cnl_tt}\ee
\be \frac{\partial }{\partial t} \cnl(t,\tw) = - \cnl(t,\tw) +
\th'(m)\cg(t,\tw)\label{eq:cnl_ttw} \ee
\be \cl(t,t) = 1-m^2(t) \label{eq:cln_tt} \ee
\be \frac{\partial }{\partial t}\cl(t,\tw) = - \cl(t,\tw)
\label{eq:cln_ttw} \ee
as shown in Appendix~\ref{app:technical} (see eqs.\eq{eq:cln_tt_app},\eq{eq:cnl_tt_app},\eq{eq:cnl_ttw_app}
and\eq{eq:cln_ttw_app}). The quantity $a$ appearing
in\eq{eq:cnl_tt} is defined as
\be a = 1 - \th'(m) = 
%1-\beta\tanh'(\beta m) =
1-\beta[1-\tanh^2(\beta m)].\label{eq:a} \ee
We will not normally write its time-dependence explicitly.

In order to complete the analysis of the dynamics of our model, we
need to find the linear response to applied external
fields. This is characterized by the response functions
\be R_{ij}(t,\tw)=\frac{\delta \lav \s_i (t)\rav}{\delta h_j^{\rm
ext}(\tw)}.\nn\ee
As for the correlation functions, we have to leading order in $N$
\bea R_{ii}&=&\rl+O(N^{-1})\label{eq:localresp}\\
R_{ij} &=& \rnl/N+O(N^{-2})\label{eq:nonlocalresp}\eea
while the leading term in the global response, of the magnetization to a uniform field, is
\be \rg = \rl+\rnl.\label{eq:rg} \ee
The evolution equations for these response functions can be expressed as
\be \frac{\partial}{\partial t} \rnl(t,\tw) =
\th'(m)[\rl(t,\tw)+\rnl(t,\tw)] - \rnl(t,\tw)
\label{eq:rnl_ttw}
\ee
\be
\frac{\partial}{\partial t} \rl(t,\tw) = -  \rl(t,\tw)
\label{eq:rln_ttw} \ee
as derived in Appendix \ref{app:technical} (see eqs.\eq{eq:rnl_ttw_app}
and\eq{eq:rln_ttw_app}).
These equations can be integrated forward in time starting from the
values of the equal-time response functions\eq{eq:rln_tt_app}:
\be \rl(t,t) = \th'(m), \qquad \label{eq:rln_tt} \rnl(t,t) =
0. \ee
The evolution equations (\ref{eq:cnl_tt}--\ref{eq:cln_ttw}) for the
correlations and (\ref{eq:rnl_ttw}--\ref{eq:rln_tt}) for the responses
contain all the relevant information about the dynamical properties of
the ferromagnetic systems.

\section{Fluctuation-dissipation ratios}
\label{sec:fdr}

Using the results obtained in the previous section, we can study in
full detail the fluctuation-dissipation ratios (FDR) for global and local relaxation functions.
Crucially, we will be able to investigate how the value of the asymptotic global FDR
depends on the initial condition. Our calculation in the infinite
system size limit will produce different values for 
global and local FDRs; this apparent breaking of the expected
local--global correspondence will
be solved and carefully explained in Section \ref{sec:corrections}.

\subsection{Global FDR}

To compute the global FDR, we first need the equal-time global
correlation $\cg(t,t)=\cl(t,t)+
\cnl(t,t)=1-m^2(t)+\cnl(t,t)$. Differentiating this expression with
respect to time and using\eq{eq:m0_t} and\eq{eq:cnl_tt} we get
\bea
\frac{\partial}{\partial t}\cg(t,t)
&=& -2m[\th(m)-m]-2a\cnl(t,t)
\nonumber\\
& &{}+{} 2(1-a)(1-m^2)
\nn\\
&=& -2a\cg(t,t)+\d \label{eq:cg_tt} \eea
with $\d=2[1-m\,\th(m)]$ and $a$ defined in\eq{eq:a}.
Equation\eq{eq:cg_tt} can be integrated explicitly to get
\be \cg(t,t)=r^2(t)\cg(0,0)+\int_0^t
dt'\,\frac{r^2(t)}{r^2(t')}\d(t') \label{eq:cg_tt_solved}\ee
where we have defined the quantity
\be r(t)=\exp\left[-\int_0^t dt'\,a(t')\right].
\label{eq:r_def} \ee
For the global two-time correlation\eq{eq:cg}, the sum of\eq{eq:cnl_ttw}
and\eq{eq:cln_ttw} gives $(\partial/\partial t)\cg(t,\tw)=-a
\cg(t,\tw)$ and after integration
\be \cg(t,\tw)=\frac{r(t)}{r(\tw)}\cg(\tw,\tw).
\label{eq:cg_explicit} \ee
To compute the corresponding global response\eq{eq:rg}, we add
equations\eq{eq:rnl_ttw} and\eq{eq:rln_ttw} to obtain
\be \frac{\partial}{\partial t}\rg(t,\tw) = -a\rg(t,\tw).\nn\ee
The solution for the global response is then
given by
\be \rg(t,\tw) = \frac{r(t)}{r(\tw)}\beta[1-\tanh^2(\beta
m(\tw))] \label{eq:rg_explicit} \ee
where we have used $\rg(t,t)=\rl(t,t)$ and\eq{eq:rln_tt}.

By combining the last two results we get an exact analytical
expression for the global FDR
\bea \xg(t,\tw)=\frac{T\rg(t,\tw)}{\frac{\partial
\cg(t,\tw)}{\partial\tw}} = \frac{T[1-a(\tw)]}
{\d(\tw)-a(\tw)\cg(\tw,\tw)} \label{eq:xg} \eea
where we have also used the fact that $\partial
r(\tw)/\partial\tw=-a(\tw)r(\tw)$. Equation\eq{eq:xg} shows
explicitly that the global FDR depends only on the earlier time
$\tw$, which is a feature often seen in simple mean-field models. From the
general expression\eq{eq:xg} we can now analyze the asymptotic FDR for
different initial conditions.

\subsubsection{Zero initial magnetization}

We consider first the standard case where the system is initially
unmagnetized. This implies $a=1-\beta$, $\d=2$ and
$r(t)=\exp(-at)$. From\eq{eq:cg_tt_solved} the equal-time correlation
is given by
\be \cg(t,t)=e^{-2at}\cg(0,0)+\left(1-e^{-2at}\right)(\d/2a).
\label{eq:cg_tt_eql} \ee
For high temperatures $T>\Tc=1$, where $a>0$, this converges
exponentially to its equilibrium value $\d/2a=1/a$; the FDR
approaches the limiting value $\xg^\infty =
T(1-a)/(2-a/a)=1$ and the system equilibrates as expected. Below
the critical temperature, on the other hand, $a$ is negative and
$\cg(t,t)$ diverges exponentially so that $\xg^\infty = 0$. At
criticality, finally, where $T=1$ and $a=0$, the equal-time
correlator grows linearly as $\cg(t,t)=\cg(0,0) + 2t$ and the FDR has
a nontrivial finite limit $\xg^\infty = 1/2$.  These
results can be summarized as follows:
\begin{equation}
\xg^\infty = \left\{
\begin{tabular}{ccl}
$0$   &\mbox{\ \ \ } & $T ~<~T_c$\\
$1/2$ && $T ~=~T_c$\\
$1$   && $T ~>~T_c$
\end{tabular}
\right.
\label{eq:beta_below_m0}
\end{equation}
These FDR values for the long-range Ising ferromagnet with zero initial
magnetization are identical to those obtained for
finite-dimensional spherical ferromagnets above their upper critical
dimension~\cite{luck1b,spherical_model_forthcoming}, as one might have
expected on physical grounds.
%
%; while the value
%$\xg^\infty=1/2$ for the global FDR is also found in the
%$d=1$-dimensional Ising model at $T=0$~\cite{luck2}.
%
%Therefore, the
%results obtained 
%magnetization are consistent with these previous studies for
%ferromagnetic spins systems.

\subsubsection{Non-zero initial magnetization}

For non-zero initial values of the magnetization, which without loss
of generality we take as positive, one again needs to distinguish
temperatures above, below and at the critical temperature. In the
first two cases, equation\eq{eq:m0_t} tells us that $m(t)$ decays
exponentially to its equilibrium value $m\eql$. This value is
$m\eql=0$ for $T>\Tc$, while for $T<\Tc$ it is the (positive) solution
of $m\eql=\th(m\eql)=\tanh(\beta m\eql)$. Along with $m$, the
quantities $a$ and $\d$
also converge quickly to $a\eql = 1-\beta(1-m\eql^2)$ and
$\d\eql=2(1-m\eql^2)$. As $a\eql$ is just the relaxation rate of
$m(t)$ to $m\eql$, it is positive both above and below the critical
temperature. From\eq{eq:cg_tt_eql} the equal-time correlator then
tends to $\d\eql/(2a\eql)$ and so the FDR approaches the value~\cite{T0note}
\be \xg^\infty = 
\frac{1-m\eql^2}{\d\eql-\d\eql/2}=1.
\label{eq:global_nonzero_m}\ee
This is as expected since the system reaches equilibrium exponentially
quickly.

The interesting case is a quench to criticality ($T_c=1$) with
non-zero initial magnetization $m(0)$. Here equation\eq{eq:m0_t}
for the magnetization yields the asymptotic power-law decay
\be m(t) = \sqrt{3/2t}\label{eq:m0_asympt} \ee
independently of initial conditions. Also $a(t)=\tanh^2(m(t))$
which at long times becomes $a(t)=3/(2t)$; as a consequence, the
function $r(t)$ from\eq{eq:r_def} scales asymptotically as
$\exp[-(3/2)\ln t]=t^{-3/2}$. For the equal-time
correlation\eq{eq:cg_tt_solved}, both the term including the
initial condition $\cg(0,0)$ and the correction arising from the
approach of $\d(t)=2[1-m(t)\tanh(m(t))]=2-3/t+\ldots$ to its limit
value are then subleading and one has at long times
$\cg(t,t)=2\int_0^t dt'\,(t/t')^{-3}=t/2$. The product
$a(t)\cg(t,t)$ thus approaches $[3/(2t)]\times(t/2)=3/4$ and the
global FDR\eq{eq:xg} tends to
\be \xg^\infty = \frac{1}{2-3/4} = \frac{4}{5}~. \nn\ee
This result can also be obtained directly from the long-time forms of
the correlation\eq{eq:cg_explicit} and response\eq{eq:rg_explicit},
\be \cg(t,\tw) = \frac{\tw}{2} \left(\frac{t}{\tw}\right)^{-3/2},
\quad \rg(t,\tw) = \left(\frac{t}{\tw}\right)^{-3/2}.\nn\ee
Again, we can summarize the results:
\begin{equation}
\xg^\infty = \left\{
\begin{tabular}{ccl}
$1$   &\mbox{\ \ \ } & $T ~<~T_c$\\
$4/5$ &              & $T ~=~T_c$\\
$1$   &              & $T ~>~T_c$
\end{tabular}
\right.
\end{equation}
The difference between these FDR values and those for the unmagnetized
case, eq.\eq{eq:beta_below_m0}, is a clear signature of the difference
in the underlying coarsening dynamics. For $T<T_c$ it is physically
obvious that the processes involved are very different: in the
magnetized case the system equilibrates exponentially quickly, whereas
for $m=0$ it ages indefinitely and equilibrium is never established.

The result that also the FDR {\em at criticality} depends on whether the
system is initially magnetized or not, on the other hand, is
highly nontrivial. Indeed, one might have expected that the difference
between the two cases becomes negligible at long times because the
magnetization decays towards zero even in the initially magnetized
scenario. Our explicit results show that this relaxation of $m(t)$
does contribute significantly to the FDR, which acquires a nontrivial
non-zero limiting value. The latter is distinct from the standard
results $X^\infty=1/2$, indicating that coarsening in the presence of
a non-zero magnetization belongs to a different dynamical universality
class from coarsening at $m=0$.

\subsection{Local FDR}

Let us now check if the results obtained for global quantities can
be reproduced from the local FDR. This is important because the
concept of a non-equilibrium temperature is based on its independence
on the choice of observable, and also because numerical work has
often focussed on the simulation of local correlation and response
functions~\cite{luck1b,chatelain,ricci}.

The equal-time values of the local correlation and response are
given in\eq{eq:cln_tt} and\eq{eq:rln_tt}. From\eq{eq:cln_ttw}
and\eq{eq:rln_ttw} the corresponding two-time quantities decay
exponentially, so that
\bea \cl(t,\tw)&=&[1-m^2(\tw)]e^{-\tau}
\label{eq:cln_explicit}
\\
\rl(t,\tw)&=&\beta[1-\tanh^2(\beta m(\tw))]e^{-\tau}
\label{eq:rln_explicit} \eea
where $\tau=t-\tw$. The FDR corresponding to the local correlation
and response follows as
\be \xln(t,\tw) = \frac{1-\tanh^2(\beta m(\tw))}
{1-m^2(\tw)-2m(\tw) \frac{\partial
m(\tw)}{\partial\tw}}\label{eq:X_local0} \ee
and again only depends on the earlier time $\tw$. If the system starts
in an unmagnetized state then $m=0$ at all times and therefore
$\xln^\infty=1$ for all temperatures. For non-zero initial values
of the magnetization and $T>\Tc$, $m(t)$ decays exponentially to
zero and so $\xln(\tw\to\infty)\to 1$. For $T<\Tc$, $m(t)$ also
decays exponentially, but to a non-zero equilibrium value $m\eql$.
Nevertheless, because $m\eql=\tanh(\beta m\eql)$,
equation\eq{eq:X_local0} implies that again $\xln(\tw\to\infty)\to
1$. At criticality, finally, inserting $m^2_0(t)=3/(2t)$
into\eq{eq:X_local0} shows that also here $\xln(\tw\to\infty)\to
1$, though the convergence is now as a power law ($\sim 1/\tw^2$)
rather than exponentially. Therefore \be \xln^\infty = 1
~~~~\mbox{for all} ~~T.\ee

In summary, the limiting FDR obtained from the local correlation and
response does not pick up any 
signature of the phase transition at $\Tc=1$, whether the system is
magnetized or not. We will see in Sec.~\ref{sec:corrections} that this
is a somewhat pathological consequence of taking $N\to\infty$ before
looking at long times, and that finite-$N$ corrections restore the
expected correspondence between local and global measurements.

\section{Finite-dimensional models for large $d$}
\label{sec:finite}

One would expect that the behaviour observed above for the
long-range model should also appear in short-range models above
their critical dimension. We therefore now extend our discussion
to the Ising model on a $d$-dimensional hypercubic lattice with
nearest neighbour (n.n.) interactions, in the limit of large $d$.
A complication in this case is that there are multiple scalings of the spin
correlation functions. For example, local correlations are $O(1)$,
those between n.n.\ spins scale as $O(1/d)$, and those between
next nearest neighbours (n.n.n.) as $O(1/d^2)$. In order to capture
the $O(1)$ contribution of these correlation functions (and the
corresponding responses) it is useful to consider the Fourier transforms
\bea \Cq(t,\tw) = \sum_l
e^{i\qv\cdot(\rv_l-\rv_j)}C_{lj}(t,\tw)\nn\\
\Rq(t,\tw) =
\sum_l e^{i\qv\cdot(\rv_l-\rv_j)}R_{lj}(t,\tw). \nn\eea
For example, the $2d$ n.n.\ spins with their correlations
$C_{i,j}(t,\tw)$ of $O(1/d)$ given an overall contribution to $\Cq(t,\tw)$
of $O(1)$; the same is true for the $O(d^2)$ n.n.n.\ spins
with their $O(1/d^2)$ correlations and so on. 

The Hamiltonian of the short-range model is, by analogy
with\eq{eq:ham},
\be H=-\frac{1}{2d}\sum_{(i,j)} \s_i\s_j - \sum_i \hext_i\s_i.
\nn\ee
In the interaction term the sum now runs over all n.n.\ pairs of
spins; the interaction strength has been chosen to get the same
critical temperature, $T_c=1$, in the limit $d\to\infty$ as in the
long-range model. The local fields are now given by $h_i =
\hext_i+ (2d)^{-1}\sum_{k} \sigma_k$ instead of eq.\eq{eq:field}, with
the sum running over all n.n.s\ of $i$. For large $d$ the field
fluctuations $\Delta h_i$ are small, of $O(d^{-1/2})$, and so one can
again linearize in $\Delta h_i$.

As for the long-range model, we proceed to study the
correlation and response functions in this model. The 
general equations\eq{eq:Cij_tt} and\eq{eq:Cij_ttw} can be used to derive
the dynamical equations for the correlations. To arrive at explicit
expressions, we need to analyze the correlations between spins and local fields. They can be expressed as
\be \lav \Delta h_i \Delta \sigma_j \rav = \frac{1}{2d}\sum_{k}
\lav \Delta \sigma_k \Delta \sigma_j\rav = \frac{1}{2d}\sum_{k}
C_{kj}\nn\ee
with Fourier transform
\be \sum_l e^{i\qv\cdot(\rv_l-\rv_j)} \lav \Delta h_l \Delta
\sigma_j \rav = (1-\omq)\Cq\nn.
\ee
Here
\be
\omq = 1 -
\frac{1}{d}\sum_{a=1}^d \cos q_a \nn\ee
and the $q_a$ are the spatial components of the wavevector
$\qv$. Using the large-$d$ expansion of the free
energy~\cite{yedidia_prb}
one can show that equilibrium correlations
involving higher powers of the field fluctuation, e.g.\
$\lav(\Delta h_i)^2\Delta \sigma_j\rav$, have Fourier transforms
which are suppressed by $O(1/d)$ (away from
criticality). Following the same reasoning as for the
long-range ferromagnet, we discard these subleading contributions.
For the magnetization, this leads back to the expected
mean-field equation of motion\eq{eq:m0_t}. The evolution of the
equal-time correlations follows by linearization and Fourier
transformation of\eq{eq:Cij_tt} as
\be \frac{\partial }{\partial t}\Cq(t,t) = -2\Cq(t,t) + 2\th'(m)
(1-\omq)\Cq(t,t) + \zz(t) \label{eq:Cq_tt} \ee
where the last term accounts for the fact that the local
correlations $C_{ii}(t,t) = 1-m^2(t)$ have a different equation of
motion from the non-local ones. One can write an explicit expression
for $\zz(t)$ but this is not helpful because it depends itself on the
$\Cq(t,t)$ which we are trying to find. Instead we first
solve\eq{eq:Cq_tt} for arbitrary
$\zz(t)$ and then determine the latter such that the local
correlations come out correctly. 
Generalizing the definitions of
$a$, eq.\eq{eq:a}, and $r$, eq.\eq{eq:r_def}, to
\bea \aq &=& 1-\th'(m)(1-\omq) \nn\\
\rq(t)&=&\exp\left[-\int_0^t
dt'\,\aq(t')\right] \nn\eea
the solution of\eq{eq:Cq_tt} reads
\be \Cq(t,t) = \rq^2(t)\Cq(0,0) + \int_0^t
dt'\frac{\rq^2(t)}{\rq^2(t')}\, \zz(t').\label{eq:Cq_tt_explicit} \ee
The function $\zz(t')$ can now be determined from the constraint
that $\dq \Cq(t,t) = C_{ii}(t,t)$, where the shorthand notation
$\dq$ indicates the integral over $\qv\in[-\pi,\pi]^d$ normalized
by $(2\pi)^d$. Integrating\eq{eq:Cq_tt_explicit} we find that
\be 1-m^2(t) = \dq \rq^2(t)\Cq(0,0) + \int_0^t \!\!dt'\!
\dq\frac{\rq^2(t)}{\rq^2(t')}\,\zz(t'). \label{eq:z_aux} \ee
For simplicity, we focus in the following on initially
uncorrelated spins, i.e.\ $\Cq(0,0)=1-m^2(0)$. Using that
\bea 
\dq e^{x(1-\omq)} &=& \left(\int_{-\pi}^\pi
\frac{dq_1}{2\pi}e^{(x/d)\cos q_1}\right)^d \nn\\
&=&
\left(1+\frac{x^2}{4d^2}+\ldots\right)^d \nn\\
&=&
1+O\left(\frac{1}{d}\right)\nn \eea
equation\eq{eq:z_aux} then simplifies for large $d$ to
\be 1-m^2(t) = e^{-2t}[1-m^2(0)] + \int_0^t dt'\,
e^{-2(t-t')}\zz(t'). \nn\ee
For $m(t)=0$ at all times this gives the constant value $\zz(t)=2$. For more
general scenarios, $\zz$ still converges to this value at long times
as long as the magnetization decays to zero, i.e.\ for $T\geq \Tc$.

For the two-time correlations, linearization of\eq{eq:Cij_ttw} in
$\Delta h_i$ yields the evolution equation $(\partial/\partial
t)\Cq(t,\tw)=-\aq\Cq(t,\tw)$ and thus
\be \Cq(t,\tw)=\frac{\rq(t)}{\rq(\tw)} \Cq(\tw,\tw). \nn\ee
The instantanteous response remains purely local as
in\eq{eq:Rij_aux}, with $\Rq(t,t)=\th'(m)$ to leading order in
$1/d$. The evolution of the two-time response is obtained as in the
long-range case, by
linearizing\eq{eq:resp} in the applied field. This gives
simply $(\partial/\partial t)\Rq(t,\tw) = -\aq \Rq(t,\tw)$, which
integrates to
\be \Rq(t,\tw) = \frac{\rq(t)}{\rq(\tw)} \th'(m(\tw)). \nn\ee
On the basis of the correlation and response functions one can
define a Fourier-component FDR. By analogy with\eq{eq:xg} this can
be simplified to
\be X_\qv(t,\tw)= \frac{T\Rq(t,\tw)}{\frac{\partial \Cq}{\partial\tw}(t,\tw)} = \frac{\th'(m(\tw))}
{\zz(\tw)-\aq(\tw)\Cq(\tw,\tw)}\ . \label{eq:Xq} \ee
The FDR for the global magnetization, which is the Fourier
component of the spins with $\qv=0$, is obtained from this expression as a
special case. Let us now concentrate on quenches to $T_c$ where we found
for the long-range model a nontrivial dependence of the results on the
initial conditions.

\subsection{Zero initial magnetization}

\begin{figure}[t]
\begin{center}
\includegraphics*[height=8.6cm,angle=-90]{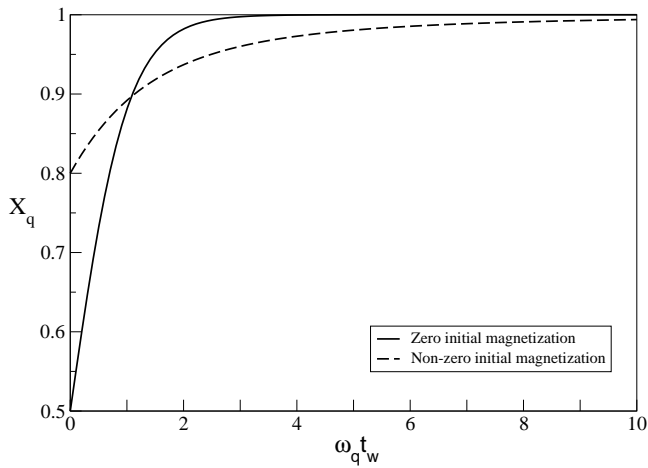}
\vskip 0.1in \caption{Dependence of the FDR on the lengthscale in the
large-$d$ short-range ferromagnet, for initial conditions with zero
(solid line, eq.\eq{eq:xq_nonmag}) and non-zero (dashed
line, eq.\eq{eq:xq_mag}) magnetization. Shown on the $x$-axis is the scaling
variable $\omq\tw\sim q^2\tw$, which is proportional to the squared
ratio of the time-dependent correlation length ($\sim \tw^{1/2}$) to
the lengthscale being probed by the chosen observable ($1/q$).
\label{fig1}}
\end{center}
\end{figure}

For zero initial magnetization, $m(t)=0$ at all times and so
$\aq(t)=\omq$ and $\rq(t)=\exp(-\omq t)$. Correspondingly, the
equal-time correlations\eq{eq:Cq_tt_explicit} simplify to
\be \Cq(\tw,\tw)=\exp(-2\omq \tw) +
\frac{1}{\omq}\left(1-e^{-2\omq \tw}\right).
\label{eq:Cq_twtw} \ee
The relevant scaling variable is clearly $w\equiv\omq \tw$, so we
will focus on the limit of long times and low ``frequencies''
$\omq$ taken such that that $w$ remains constant. Since
$\omq=\qv^2/(2d)$ for small $\omq$, the scaling $\omq\sim\tw^{-1}$ reflects the
growing lengthscale $1/q\sim \tw^{1/2}$ of the correlations as the
system coarsens. The first term in\eq{eq:Cq_twtw}, which arises
from the decay of initial correlations, then becomes negligible
compared to the second and we get
\bea
 X_\qv(t,\tw) &\simeq& \frac{1}{2-\omq [1-\exp(-2\omq \tw)]/\omq}
\\
&=& \frac{1}{1+\exp(-2\omq \tw)}\ . \label{eq:xq_nonmag}\eea
For lengthscales short compared to the time-dependent correlation
length, where $\omq\tw\gg 1$, $X_\qv$ becomes equal to unity as
expected because of effective equilibration on such short scales.
For much larger lengthscales ($\omq\tw\ll 1$), and in
particular in the limit $\omq\to 0$ which gives the FDR for the
magnetization, $X_\qv$ approaches $1/2$. These two limits
can be seen in Fig.~\ref{fig1} and are consistent with
the results for the long-range ferromagnet, but here we see that 
in addition one can interpolate 
smoothly between the two limits by varying the lengthscale
considered. Similar behaviour is also found in the Ising
chain~\cite{soll2}, i.e.\ for $d=1$, and in the spherical
model~\cite{spherical_model_forthcoming}. Given that our calculation
is based on a linearization in the local field fluctuations, it is not
surprising that the result\eq{eq:xq_nonmag} can also be obtained from
a Gaussian field theory~\cite{CalGam02}.

\subsection{Non-zero initial magnetization}

For non-zero initial magnetization and at criticality, the
magnetization again decays according to\eq{eq:m0_asympt}.  One then
has $\th'(m(\tw))=1-3/(2\tw)$ to leading order for large times, and
consequently $\aq(\tw)=1-[1-3/(2\tw)](1-\omq)$. This gives
\bea \rq(t) &=& \exp\left[-\int_0^t dt'\,\aq(t')\right]\nn\\
&=& e^{-t +
[t-(3/2)\ln t](1-\omq)} \nn\\
&=& t^{-(3/2)(1-\omq)}e^{-\omq t}\nn\eea
and for long times, the equal-time correlations\eq{eq:Cq_tt_explicit}
follow as
\bea \Cq(\tw,\tw) &=& \tw^{-3(1-\omq)}e^{-2\omq \tw}
\nn\\
& & {}+{} \int_0^{\tw}
dt' \left(\frac{t'}{\tw}\right)^{3(1-\omq)} e^{-2\omq(\tw-t')}
\zz(t').\nonumber \eea
For $\tw\to\infty$ at fixed $w=\omq\tw$, the first term and the
approach of $\zz(t')$ to its limit $\zz=2$ give only subleading
corrections and one gets asymptotically
\bea \Cq(\tw,\tw) &=& \frac{1}{\omq} \scc(\omq\tw)\nn\\
\scc(w) &=& 2w \int_0^1 dy\, y^3 e^{-2w(1-y)}. \nn\eea
The Fourier-component FDR\eq{eq:Xq} therefore becomes
\bea X^{-1}_\qv(t,\tw) &=& 2-\left\{
1-[1-3/(2\tw)](1-\omq)\right\} 
\nonumber\\
& &~~~~~\times\omq^{-1}\scc(\omq\tw)\nn
\\
&=& 2- [3/(2w)+1]\scc(w) \label{eq:xq_mag} \eea
where we have neglected all terms that are subleading ($\sim
1/\tw$) for long times. For short lengthscales, $w\gg 1$, one sees
easily that $\scc \to 1$ and so $X_\qv\to 1$, demonstrating the expected
equilibration. On the other hand, for lengthscales much larger
than the time-dependent correlation length, i.e.\ $w\ll 1$, one
has $\scc=w/2$ and so $X_\qv\to (2-3/4)^{-1}=4/5$. This applies in
particular to the FDR for the magnetization ($\qv=0$) and so is
consistent with the result found above for the long-range model.
Again there is a smooth lengthscale-dependence of the FDR that
interpolates between local equilibrium and the nontrivial FDR,
$X^\infty=4/5$, for large lengthscales; see the dashed line in
Fig.~\ref{fig1}. This is rather reassuring:
it tells us that there is nothing special about the magnetization,
i.e.\ $\qv=0$, even though in a magnetized system this is the only
Fourier component that has a non-zero average. In physical terms,
the FDR for the magnetization can also be observed by looking at
lengthscales that are much smaller than the system size, as long
as they are large compared to the time-dependent correlation
length.

We have only discussed the FDR for pure Fourier components here.
However, if one considers more general correlations, say of spins
across some finite range, one simply gets a mixture of the Fourier
component FDRs. The resulting $X(t,\tw)$ will be a mixture of all
$X_\qv(t,\tw)$ and interpolate between
$X(\tw,\tw)=1$ at equal times and a nontrivial asymptotic value
$X^\infty$ in the limit $t\gg\tw$ of well separated times.
Following the reasoning in the one-dimensional case~\cite{soll2}, one can
show that this asymptotic FDR is always identical to the FDR for
the longest wavelength, i.e.\ $X_{\qv=0}$, because the limit $t\gg\tw$
suppresses the contributions from all non-zero wavevectors. In particular, this
means that {\em all} observables that are linear in the spins,
including local correlation and response, will give $X^\infty=1/2$
for critical coarsening at zero magnetization, and $X^\infty=4/5$
for the magnetized case. (Similarly to the long-range case discussed in
the next section one can show that these asymptotic FDR values
would only be observed for rather long time differences $t-\tw$, of
order $2^d$ or larger, while for shorter time differences
apparent equilibrium behaviour is obtained.)

\section{Finite-size corrections for the long-range ferromagnet}
\label{sec:corrections}

In the short-range model just discussed, local and global observables
give the same limiting FDR $X^\infty$. But in the long-range model of
Sec.~\ref{sec:fdr}, this correspondence appears to be broken because the
local correlation and response functions do not pick up any
non-equilibrium effects. To analyze the origin of this discrepancy, we
now study the $1/N$ corrections to $\cl$\eq{eq:local} and
$\rl$\eq{eq:localresp}. The calculations are sketched in
Appendix~\ref{app:corrections} and we only give the main results here.

Let us first consider again the unmagnetized case at high
temperatures, $T>\Tc$. Keeping terms up to $O(N^{-1})$, we find for
the local correlation
\be \wcl(t,\tw)=e^{-\tau}\left[1-\frac{(1-\beta)^{-1} +
\beta\tau}{N}\right] +\frac{e^{-(1-\beta)\tau}}{N(1-\beta)}
\label{eq:clno_para} \ee
where $\tau=t-\tw$ as before. The second term becomes dominant over the
first for time differences $\tau \approx T\ln N$, i.e.\ on a timescale
that grows only logarithmically with the system size. (The
$1/N$-corrections in the square brackets can always be neglected;
they would become relevant only for $\tau\sim N$, but by then the
second term is exponentially larger than the first.) For a
finite-size system the correction term therefore dominates the
long-time dynamics, causing the decay rate of the correlation to
slow from 1 to $1-\beta$ at $\tau\approx T \ln N$.  The corresponding
local response is related by FDT to the correlation; this is
as expected because for the long times considered here the system is
in equilibrium. A similar argument indicates that the above
long-time results for $T>\Tc$ remain unchanged if the initial
magnetization is non-zero.

Next we analyse the non-equilibrium dynamics at criticality
($T=T_c=1$) starting from zero magnetization; here we would hope
to retrieve from the $O(1/N)$ correction terms a long-time FDR of
$\xl^\infty=1/2$. One finds
\be \wcl(t,\tw) = e^{-\tau} + \frac{1}{N}\left[
\cg(\tw,\tw)\left(1-e^{-\tau}\right) -  \tau\,
e^{-\tau}\right]. \label{eq:clo_crit0} \ee
Using the fact that $\cg(\tw,\tw)=2\tw+\cg(0,0)$ at criticality, the
$1/N$-expansion now breaks down for $\tw \sim N$, where the
correction term becomes $O(1)$ rather than $O(N^{-1})$ as the
expansion assumes. For smaller values of $\tw$ the expansion remains
valid, however, and the $\tw$-derivative of the correlation
becomes
\bea \frac{\partial}{\partial \tw}\wcl(t,\tw) &=& e^{-\tau} +
\frac{2}{N}\left(1-e^{-\tau}\right) \nonumber\\
& &{}-{}
\frac{1}{N}e^{-\tau}[\cg(\tw,\tw) + 1+\tau]. \nn\eea
The third term can be neglected in the reliable regime $\tw\ll N$,
but the second one again becomes dominant over the leading
contribution for $\tau\approx \ln N$; for $\tau-\ln N\gg 1$, one then has
$(\partial/\partial \tw)\cl(t,\tw) =2/N$. The corresponding local
response reads
\be
\wrl(t,\tw)=e^{-\tau}\left\{1-\frac{1}{N}\left[\cg(\tw,\tw)+1+\tau\right]\right\}
+ \frac{1}{N}\ . \label{eq:rlno_crit} \nn\ee
The second term becomes dominant over the first for $\tau\approx \ln
N$, while the $1/N$-corrections in the first term can always be
neglected for $\tw\ll N$. For $\tau-\ln N\gg 1$ the corrected FDR is therefore
$\wxl(t,\tw)=(1/N)/(2/N)=1/2$. To summarize, for systems that are old
($\tw\gg 1$) 
but not yet equilibrated ($\tw\ll N$), the FDR as a function of
the time difference $\tau$ crosses over from the equilibrium value
$\wxl=1$ to the non-equilibrium value $\wxl^\infty=1/2$ on a
timescale $\tau\approx\ln N$. In any finite-size system the limiting
value of the {\em local} FDR therefore agrees with the {\em
global} one, just as the local--global correspondence leads one to
expect. If the limit $N\to\infty$ is taken before the long time
limit, as we did in section~\ref{sec:infinite_N}, then one
implicitly discards the non-equilibrium regime. This is what leads
to the apparent breaking of the correspondence with the global results.

We now compare with the corresponding results for the
non-equilibrium dynamics at criticality starting from non-zero initial
magnetization.
Evaluating the $1/N$-correction terms in the regime $\tau\gg 1$
where they are potentially relevant, we find
\bea \wcl(t,\tw) &=& e^{-\tau} +  \frac{\tw}{2N}
\left(\frac{t}{\tw}\right)^{-3/2}\label{eq:clno_crit_m}
\\
\wrl(t,\tw) &=& e^{-\tau} +
\frac{1}{N}\left(\frac{t}{\tw}\right)^{-3/2}.
\label{eq:rlno_crit_m} \eea
As in the unmagnetized case, the correction terms become significant for
$\tau\approx\ln N$. For larger time differences the local quantities
become proportional to the global ones; as a consequence, the
corrected local FDR $\wxl$ crosses over from 1 to the global FDR
$\wxl^\infty=\xg^\infty=4/5$. Our main conclusion of this section is,
therefore, that in any finite-size system the local--global
correspondence is preserved.

We note as an aside that the relevant
scaling variable for the above crossovers in $\wxl$ is $N\exp(-\tau)$.
Plots of $\wxl$ versus $N\exp(-\tau)$ would look similar to those in
Fig.~\ref{fig1} for both the magnetized and unmagnetized cases, though
one has to bear in mind that somewhat different quantities are being
plotted: Fig.~\ref{fig1} refers to the lengthscale-dependence of the
limiting FDR, whereas here we have a fixed short lengthscale (local
correlation and response) and are looking at the system-size dependent
crossover in time of the FDR to its limiting value.

We have not considered the $1/N$-corrections at $T<\Tc$ in this
section because it turns out -- consistent with the fast equilibration
when starting from non-zero magnetization -- that here the
$1/N$-expansion breaks down already for system ages $\tw$ of order
$\ln N$.

\section{Summary and discussion}
\label{sec:summary}

In this paper we have solved analytically the non-equilibrium dynamics
of the long-range Ising ferromagnet with Glauber dynamics, initially in the
thermodynamic limit and then including also the leading finite-size
corrections; our focus was on the correlation and response functions
and the associated fluctuation-dissipation ratio (FDR).  We have also
analysed the corresponding short-range model in the limit of large
dimension, which 
provides useful additional insights into the
lengthscale-dependence of the FDR.

Our main result is that different nontrivial values of the limiting
FDR $X^\infty$ can result depending on the initial conditions. In
particular, for quenches to the critical temperature we find that
$X^\infty=1/2$ in the standard scenario where the system is initially
unmagnetized, while $X^\infty=4/5$ if the initial magnetization is
non-zero. We are not aware of any previous observations of a
nontrivial (non-zero) 
value of $X^\infty$ arising from initial conditions other than those
traditionally considered; earlier studies~\cite{PicHen02,henkel} of
strongly correlated initial conditions had always found either the
standard value or $X^\infty=0$.

Our findings show that critical coarsening processes are fundamentally
different depending on whether the system is magnetized or not, and
the two cases must be considered as belonging to different dynamical
universality classes. One would have certainly expected such a
distinction {\em below} $T_c$, where in the magnetized case the system
equilibrates rapidly. Our finding that the difference
persists even {\em at} $T_c$ is much less obvious, seeing as even in the
initially magnetized case the magnetization does decay towards zero at
long times. The limiting FDR thus turns out to be a useful probe for
distinguishing different classes of non-equilibrium dynamics. Of
course, the differences in the FDR also imply that the associated
effective temperatures $T/X^\infty$ differ between the magnetized
and unmagnetized cases; this emphasises that the same system can have
different non-equilibrium effective temperatures depending on its
initial preparation, which then reflect the physical differences in the
ensuing non-equilibrium dynamics.

One may ask whether our results are peculiar to the relatively simple
scenarios that we have considered. This is not so: calculations in the
spherical ferromagnet quenched to
criticality~\cite{spherical_model_forthcoming} give exactly the same
limiting FDR $X^\infty=4/5$ for the magnetized case, in all dimensions
$d>4$. To understand where this particular value comes from, one can
write down a phenomenological Langevin equation for the evolution of
the fluctuating magnetization,
\be \frac{dm}{dt}= -m^n+h+\xi \label{eq:langevin}
\ee
where $h$ is an external magnetic field and $\xi$ is white noise with
variance $O(1/N)$. In the case $m=0$, because fluctuations around this
value will be of $O(N^{-1/2})$, the nonlinear term on the right-hand
side of eq.\eq{eq:langevin} can be neglected. One thus recovers simple
diffusive dynamics independently of $n$ (as long as $n>1$) which
results in the familiar value $X^\infty=1/2$ for the limiting FDR.  On
the other hand, if the magnetization is initially non-zero then
eq.\eq{eq:langevin} predicts that its average value decays as $\langle
m\rangle\sim t^{-1/(n-1)}$ when there is no external field, whereas
for $h\neq 0$ it approaches $\langle m\rangle \sim h^{1/n}$. Comparing
with the standard scalings $\langle m\rangle\sim t^{-\beta/(\nu z)}$
and $\langle m\rangle\sim h^{1/\delta}$, respectively, one sees that
we require $n=\delta$ and $1/(n-1)=\beta/(\nu z)$; these two choices
for $n$ are consistent with each other only if the mean-field relation
$z=2-\eta$ holds. We can then go ahead and, by linearizing
eq.\eq{eq:langevin} in the small deviations of $m$ from its average
$\langle m\rangle$, calculate its correlation and response
function. This simple calculation gives for the limiting FDR
$X^\infty=(3n-1)/(4n-2)=(2\beta+3\nu z)/(2\beta+4\nu z)$.  Inserting
the mean-field exponents $z=2$, $\beta=1/2$ and $\nu=1/2$, which imply
$n=3$, then leads to $X^\infty=4/5$.

Summarizing, the phenomenological Langevin equation\eq{eq:langevin}
predicts two different results for the limiting FDR
of mean-field ferromagnets, depending on the initial condition:
\begin{equation}
\xg^\infty = \left\{
\begin{tabular}{ll}
1/2 &\mbox{\ \ if } $m=0$\\
4/5 &\mbox{\ \ if } $m\neq 0$
\end{tabular}
\right.
\nn
\end{equation}
These are precisely the values that we found in our explicit calculations for
the long-range and high-dimensional short-range models. One may then
wonder whether this phenomenological approach can also be used to
predict FDR values 
below the upper critical dimension. The spherical model in $d<4$, for
example, has non-mean field exponents but still satisfies $z=2-\eta$,
so that the Langevin description is at least internally consistent. However,
explicit calculations~\cite{spherical_model_forthcoming} show that
the values of the FDR it predicts are incorrect. The Langevin
dynamics\eq{eq:langevin} is therefore too simple to capture the
full physics away from mean-field. It remains an open question whether
appropriately generalized phenomenological descriptions can be used to
rationalize FDR values in non-mean-field systems.

We have also investigated in this paper the dependence of the limiting
FDR on the observable considered. If the FDR and any associated
effective temperature are physically meaningful, one would hope that
e.g.\ local and global observables would lead to the same limiting FDRs.
In the short-range model we showed that this is indeed the case, because
the behaviour of both types of observables becomes dominated by the
slowest, longest-wavelength Fourier modes ($q\approx 0$) in the
limit of long times. The FDR for the Fourier modes themselves showed
the expected crossover between the values $X_\qv^\infty=1/2$ and $4/5$
for lengthscales larger than the
time-dependent correlation length ($q\ll \tw^{-1/2}$) and
$X_\qv^\infty=1$ for shorter, equilibrated, lengthscales.

In the long-range model, we found that great care is needed when
computing local FDRs because the limits $N\to\infty$ and
$t-\tw\to\infty$ do not commute. If the thermodynamic limit $N\to\infty$
is taken before the long-time limit (as we did in
section~\ref{sec:infinite_N}) one gets FDR values which are different
from those obtained for global quantities because the non-equilibrium
regime is effectively excluded. To find physically meaningful results,
one has to take the long-time limit before the thermodynamic one; this
then requires that the system size is kept large but finite as in
section~\ref{sec:corrections}. With this, the expected correspondence
between local and global FDRs is recovered. These findings are not
only of theoretical interest but also have
two implications for numerical studies. First, if in long-range models
one uses local observables to measure non-equilibrium FDRs,
simulations out to very long time differences will be required to
obtain meaningful results that can reveal non-equilibrium
effects. Second, the fact that for global
observables we could go directly to the infinite-system size limit
underlines the general message~\cite{soll2,comm} that such global
quantities are much more robust tools for detecting FDT violations
than their local counterparts.

Our main finding that magnetized and unmagnetized coarsening processes
at criticality can belong to different dynamical universality classes
clearly deserves wider study in future work. Calculations for the
spherical model~\cite{spherical_model_forthcoming} show that this
distinction also holds true below the upper critical dimension $d=4$,
and provide explicit (and surprisingly nontrivial) predictions for the
resulting FDR values. It would be interesting to complement this with
simulation studies of Ising models in $d=2$ and $d=3$. Field-theoretic
renormalization-group calculations~\cite{CalGamreview} might also be
possible for the $O(n)$ and $n$-vector models. For $n=1$ these reduce
to the Ising universality class, while for $n\to\infty$ one would
expect to recover the spherical
model results; knowledge of the detailed dependence of the FDR values
on the order 
parameter dimensionality $n$ as one interpolates between these two
extreme values should help to round out the physical
picture further. After the present work was completed we became aware
that a first step in this direction has very recently been
taken by the authors of Ref.~\cite{FedTri}, who calculated the FDR for
the $n$-vector model with a magnetized initial state within an
$\epsilon$-expansion around $d=2$.

Finally, there is the possibility that there might be yet
other initial conditions which give rise to distinct and nontrivial
values of the limiting FDR. This does not seem likely, given that the
obvious candidate case of strongly correlated but unmagnetized
configurations gives $X^\infty=0$ and can thus be excluded. Nevertheless,
a complete characterization of possible classes of
non-equilibrium coarsening induced by different initial conditions
certainly remains to be achieved.

\appendix

\section{Dynamical equations for the long-range ferromagnet}
\label{app:technical}

In this appendix we outline the explicit calculation of the dynamical
equations for the correlations and responses. The global correlation function is decomposed to leading order in the system size as
\bea \frac{1}{N}\sum_{ij}C_{ij} &=& \left[\cl+O(N^{-1})\right]
\nonumber\\
& &{}+{}(N-1)\left[\frac{\cnl}{N}+O(N^{-2})\right] \nn\\
&=&
\cl+\cnl+O(N^{-1})\nn\\
&=&\cg+O(N^{-1}) \label{eq:Cij}\eea
using eqs.~(\ref{eq:local}--\ref{eq:cg}).
At equal times, the local correlations are trivially
\be C_{ii}(t,t)=\cl(t,t) = 1-m^2(t) \label{eq:cln_tt_app} \ee
while for the non-local correlations\eq{eq:Cij_tt} implies to
leading order
\bea \!\!\!\!\!\!\!\!\!\frac{1}{N}\frac{\partial}{\partial t}\cnl(t,t) &=&
-\frac{2}{N}\cnl(t,t) + \th'(m) 
\nonumber\\
&&\times(\lav \Delta h_i \Delta \s_j \rav
+ \lav \Delta \s_i \Delta h_j \rav)
\nn
\\
&=& -\frac{2}{N}\cnl(t,t) + \frac{2}{N}\th'(m)\cg(t,t).
\label{eq:cnl_temp} \eea
In the second line we have used that, for $i\neq j$,
\bea \lav\Delta h_i \Delta \s_j\rav &=& \frac{1}{N-1} \sum_{k\neq
i} \lav
\Delta \s_k \Delta \s_j \rav \nn\\
&=& \frac{\cl(t,t) + O(N^{-1})}{N-1}
\nonumber\\
& &{}+{}
\frac{N-2}{N-1}\left[\frac{\cnl(t,t)}{N}+O(N^{-2})\right]
\nn\\
& = & \frac{\cg(t,t)}{N} + O(N^{-2}). \label{eq:h_i_sigma_j}
\eea
Defining the quantity $a$ as in\eq{eq:a}
and bearing in mind that $\cg(t,t)=1-m^2(t)+\cnl(t,t)$, we can
then write the evolution equation\eq{eq:cnl_temp} for the
non-local equal-time correlations as
\be \frac{\partial}{\partial t}\cnl(t,t) = -2a\cnl(t,t) +
2\th'(m)(1-m^2) \label{eq:cnl_tt_app}. \nn\ee
For the two-time correlations we get similarly
from\eq{eq:Cij_ttw}
\bea \frac{\partial}{\partial t} C_{ij}(t,\tw) &=& - C_{ij}(t,\tw)
+\th'(m)
\label{eq:nonloc_aux}\\
&&\times\lav \Delta h_i(t)\Delta \s_j(\tw)\rav + O(N^{-2})
\nonumber\eea
which gives for the leading order of the non-local terms the
dynamical evolution
\be \frac{\partial }{\partial t} \cnl(t,\tw) = - \cnl(t,\tw) +
\th'(m)\cg(t,\tw). \label{eq:cnl_ttw_app} \ee
For local correlations, on the other hand, the second term
in\eq{eq:nonloc_aux} is subleading so that to leading order
\be \frac{\partial }{\partial t}\cl(t,\tw) = - \cl(t,\tw).
\label{eq:cln_ttw_app} \ee

Finally we want the dynamical equations for the linear response
functions $R_{ij}(t,\tw) = \delta\lav \s_i(t)\rav/\delta
\hext_j(\tw)$. We assume that the field is applied to site $j=1$ and, as for the correlation functions, use the
appropriate scalings\eq{eq:localresp} and\eq{eq:nonlocalresp}.
Setting $\hext_1\equiv h$ and using\eq{eq:si_tt} we can write
\be \frac{\partial }{\partial t}\lav \s_i\rav = \lav \th(m+\Delta
h_i+h\delta\lav h_i\rav)\rav - \lav \s_i\rav.\label{eq:resp}
\ee
Here $m$ refers to the value of the magnetization for the
unperturbed system, while
\be \delta\lav h_i\rav = \frac{1}{N-1}\sum_{k\neq
i}R_{k1}(t,\tw) \label{eq:resp_aux} \ee
is the response function for the average value of the local field
at site $i$. If we expand again in powers of $\Delta h_i$, the
first nontrivial term is proportional to $\lav (\Delta h_i)^2\rav$
and does not contribute to leading order. Writing $\lav \s_i\rav =
m+hR_{i1}(t,\tw)$, equation\eq{eq:resp} thus becomes
\be \frac{\partial m}{\partial t} + h \frac{\partial}{\partial t}
R_{i1}(t,\tw) = \th(m+h\delta\lav h_i\rav) - m - h
R_{i1}(t,\tw). \nn\ee
Expanding to linear order in $h$ then gives back at $O(h^0)$ the
expected dynamical equation\eq{eq:m0_t} for $m$, while at $O(h)$
one gets
\be \frac{\partial}{\partial t} R_{i1}(t,\tw) = \th'(m) \delta\lav
h_i\rav - R_{i1}(t,\tw). \nn\ee
In the non-local case ($i\neq 1$), where $\delta\lav h_i\rav=
(\rl+\rnl)/N+O(N^{-2})$ from\eq{eq:resp_aux}
and (\ref{eq:localresp},\ref{eq:nonlocalresp}), this gives to leading
order
\be \frac{\partial}{\partial t} \rnl(t,\tw) =
\th'(m)[\rl(t,\tw)+\rnl(t,\tw)] - \rnl(t,\tw).
\label{eq:rnl_ttw_app} \ee
For the local case $i=1$, on the other hand, the term proportional
to $\delta\lav h_i\rav$ is subleading and one gets simply
\be \frac{\partial}{\partial t} \rl(t,\tw) = -  \rl(t,\tw).
\label{eq:rln_ttw_app} \ee
These equations can be integrated forward in time once the
instantaneous response is known. The latter is purely local, as
one sees from
\bea R_{ij}(t,t) &=& \left.\frac{\partial}{\partial \hext_j} \lav
\th(\hext_i+m+\Delta h_i) - \sigma_i\rav
\right|_{\hext=0} 
\nn\\
&=& \delta_{ij} \lav\th'(m+\Delta h_i)\rav. \label{eq:Rij_aux}
\eea
Thus, to leading order
\be \rl(t,t) = \th'(m), \qquad  \rnl(t,t) =
0. \label{eq:rln_tt_app}\ee

\section{$1/N$-corrections for the long-range
ferromagnet}
\label{app:corrections}

To calculate the $1/N$-corrections to the local correlation and
response we expand $C_{ii} = \cln + \clo/N + O(N^{-2}) =
\wcl+ O(N^{-2})$ and
$R_{ii}=\rln+\rlo/N+O(N^{-2})=\wrl+ O(N^{-2})$; the
magnetization, which enters $\clo$, is similarly expanded as
$m=m_0+m_1/N + O(N^{-2})$. The quantities $\cln$, $\rln$ and $m_0$
are then the leading order values calculated in
Sec.~\ref{sec:infinite_N}. Specifically, the global correlation
and response are to leading order $\cg=\cln+\cnl$ and
$\rg=\rln+\rnl$ as before; we will not try to calculate
$1/N$-corrections to these global quantities because these would
require the subleading corrections to the non-local terms and thus
an accurate treatment of quantities of $O(N^{-2})$.

In\eq{t_i_expansion} we now need to keep the quadratic term in
$\Delta h_i$. This gives for the magnetization:
\bea \frac{\partial m}{\partial t} &=& -m + \lav t_i \rav\nn\\
&=& -m +
\th(m) + \frac{\lav(\Delta h_i)^2\rav}{2} \th''(m) + O(N^{-2}).
\nn\eea
From the definition of $h_i$, $\lav(\Delta h_i)^2\rav = \lav
\Delta h_i \s_j\rav$ for $j\neq i$, a quantity that we worked
out in\eq{eq:h_i_sigma_j}. Expanding all quantities
in the previous equation to order $O(N^{-1})$ then gives for the
magnetization correction
\be \frac{\partial m_1}{\partial t} = -am_1 +
\frac{1}{2}\cg(t,t) \th''(m_0). %\label{eq:m1_t} 
\nn\ee
This can be integrated but we will not give the explicit result here
since it is not needed below.
%
%%If we take the initial magnetization to be independent of $N$, so
%that $m_0(0)=m(0)$ and $m_1(0)=0$, this can be integrated to give
%the general solution:
%%
%\be m_1(t) = \half \int_0^t dt'\,\frac{r(t)}{r(t')}
%\cg(t',t')\th''(m_0(t'))~~~~. \label{eq:m1_explicit} \ee
%%

For the correlations, by expanding eq.\eq{eq:cln_tt} to $O(1/N)$ we arrive at
\be \clo(t,t) = -2m_0(t) m_1(t). \label{eq:clo_tt} \ee
For $t\neq \tw$ we can use\eq{eq:nonloc_aux} with $i=j$. Bearing
in mind that $\lav h_i(t)\Delta \s_i(\tw)\rav =
\cnl(t,\tw)/N+O(N^{-2})$, the $O(N^{-1})$ terms give
\be \frac{\partial }{\partial t}\clo(t,\tw) = - \clo(t,\tw) +
\th'(m_0)\cnl(t,\tw)
%\label{eq:clo_ttw}
\nn\ee
which integrates to
\bea \clo(t,\tw) &=& e^{-(t-\tw)}\clo(\tw,\tw)
\label{eq:clo_explicit} \\
&&{}+{} \int_{\tw}^t
dt'\,e^{-(t-t')}\th'(m_0(t'))\cnl(t',\tw).
\nonumber
\eea

For the response, keeping the $(\Delta h_i)^2$ term in the
expansion of\eq{eq:Rij_aux} gives
\be R_{ij}(t,t) = \delta_{ij} \left[\th'(m) +
\frac{\cg(t,t)}{2N}\th'''(m)\right] + O(N^{-2}) \nn\ee
and the $O(N^{-1})$ terms show that the correction to the local
instantaneous response is
\be \rlo(t,t) = m_1\th''(m_0)+\half \cg(t,t) \th'''(m_0).
\label{eq:rlo_tt} \ee
For the two-time response, equation\eq{eq:resp} with the $(\Delta
h_i)^2$ term retained becomes
\bea
\frac{\partial m}{\partial t}
+ h \frac{\partial}{\partial t} R_{i1}(t,\tw)
&=&
\th(m+h\delta\lav h_i\rav) +\half \lav (\Delta h_i)^2\rav \nonumber\\
& &
\times\th''(m+h\delta\lav h_i\rav) \label{eq:resp2} 
\\
&&{}-{} m - h R_{i1}(t,\tw) + O(N^{-2}). \nonumber
\eea
To make progress we assume that the change of the variance
$\lav(\Delta h_i)^2\rav$ caused by the field $h$ is $O(h/N^2)$
rather than $O(h/N)$ and can therefore be neglected. This can be
made plausible by looking at the instantaneous response: the
variance $\lav (\Delta \s_1)^2\rav$ is changed by an amount of
$O(h)$, while changes in all other covariances $\lav \Delta \s_i
\Delta \s_j\rav$ vanish. Thus $\lav (\Delta h_i)^2\rav$ is indeed
perturbed by a negligible amount $O(h/N^2)$, and one expects the
response at later times to get no larger. We can therefore replace
$\lav(\Delta h_i)^2\rav$ by $\cg(t,t)/N$ as before and regard it
as $h$-independent to the order in $1/N$ we are retaining. The
$O(h)$ terms in\eq{eq:resp2} then yield
\bea \frac{\partial}{\partial t} R_{i1}(t,\tw) &=&
\left[\th'(m) +
\frac{\cg(t,t)}{2N} \th'''(m)\right] \delta\lav h_i\rav \nonumber\\
&&{}-{}R_{i1}(t,\tw) + O(N^{-2}). \nn\eea
For the local case of interest here, $i=1$, one has $\delta\lav
h_i\rav=\rnl/N+O(N^{-2})$ from\eq{eq:resp_aux}. Therefore:
\bea
&&\!\!\!\!\!\!\!\!\frac{\partial}{\partial t} \rln(t,\tw)
+\frac{1}{N}\frac{\partial}{\partial t} \rlo(t,\tw) =
\th'(m_0)\frac{1}{N}\rnl(t,\tw)
\nn\\
&&~~~~{}-{}\rln(t,\tw) -
\frac{1}{N}\rlo(t,\tw)
 + O(N^{-2}).
%\label{eq:resp4}
\eea
The $O(N^{-1})$ terms give the desired equation of motion for the
response correction
\be \frac{\partial}{\partial t} \rlo(t,\tw) =  - \rlo(t,\tw) +
\th'(m_0) \rnl(t,\tw) %\label{eq:rlo_ttw}
\nn\ee
which integrates to
\bea \rlo(t,\tw) &=& e^{-(t-\tw)}\rlo(\tw,\tw) \label{eq:rlo_explicit} \\
&&{}+{} \int_{\tw}^t
dt'\,e^{-(t-t')}\th'(m_0(t'))\rnl(t',\tw).
\nonumber
\eea
%
%The non-local quantities appearing in the integrals can in both cases
%be obtained by taking differences of the global and local results derived
%previously, i.e.\ $\cnl=\cg-\cln$ and $\rnl=\rg-\rln$.

In evaluating the above general predictions we start with the
unmagnetized case, where $m_0(t)=m_1(t)=0$ at all times.
From\eq{eq:clo_tt} the local equal-time correlation receives no
correction, i.e.\ $\clo(t,t)=0$, while for the response
$\rlo(t,t)=-\beta^3\cg(t,t)$ from\eq{eq:rlo_tt}. For $T>\Tc$, we
have seen that $\cg(t,t)$ approaches its equilibrium value
$1/a=1/(1-\beta)$ exponentially. With $r(t)=\exp(-at)$
and\eq{eq:cg_explicit} the global two-time correlation is then
$\cg(t,\tw)=a^{-1}\exp(-a\tau)$, while its local analogue is given
by\eq{eq:cln_explicit} as $\cln(t,\tw)=\exp(-\tau)$. Thus the
correction $\clo(t,\tw)$ to the local correlation is for long times
\bea
\beta\int_{\tw}^t dt'\,e^{-(t-t')}\left[a^{-1}e^{-a(t'-\tw)} -
e^{-(t'-\tw)}\right]~~~~~~\label{eq:clo_para}\\
= (1-\beta)^{-1} e^{-(1-\beta)\tau} -
e^{-\tau}\left[(1-\beta)^{-1} + \beta\tau\right]\nn
\eea
Combining this with the leading order term gives the
result\eq{eq:clno_para} discussed in the main text. To work out
the corresponding correction to the response one notes
from\eq{eq:rlo_tt} that the equal-time value $\rlo(t,t)=-\beta^3
\cg(t,t)$ converges exponentially to $-\beta^3/a$. Also,
the global response is $\rg(t,\tw)=\beta \exp(-a\tau)$
from\eq{eq:rg_explicit} and the local one
$\rg(t,\tw)=\beta\exp(-\tau)$ according to\eq{eq:rln_explicit}. By
inserting these results into\eq{eq:rlo_explicit} one finds
\bea
\rlo(t,\tw)
%&=& 
%-\frac{\beta^3}{a}e^{-(t-\tw)}
%+ \beta \int_{\tw}^t dt'\,e^{-(t-t')}\left[\beta e^{-a(t'-\tw)} -
%\beta e^{-(t'-\tw)}\right]\nonumber\\
&=& \beta e^{-(1-\beta)\tau} \nonumber\\
&&{}-{}e^{-\tau}\left(\frac{\beta^3}{1-\beta} + \beta +
\beta^2\tau\right)\nn\eea
and it is easy to check that this is related by FDT to the
correlation correction\eq{eq:clo_para} as it should be.

Next consider the out-of-equilibrium dynamics at criticality
($T=1$) starting from zero magnetization. For the correlation
correction we use that $\cg(t,\tw)=\cg(\tw,\tw)$ (a consequence
of\eq{eq:cg_explicit} together with $r(t)=1$), while
$\cln(t,\tw)=\exp(-\tau)$ as before. Then\eq{eq:clo_explicit}
results in
\bea
\clo(t,\tw) &=& \int_{\tw}^t
dt'\,e^{-(t-t')}\left[\cg(\tw,\tw)-e^{-(t'-\tw)}\right]\nonumber\\
&=& \cg(\tw,\tw)\left(1-e^{-\tau}\right) -  \tau\, e^{-\tau}\nn
\eea
and adding the leading order term gives\eq{eq:clo_crit0}. The
calculation for the response correction proceeds similarly, with
$\rg(t,\tw)=1$ from\eq{eq:rg_explicit} and
$\rl(t,\tw)=\exp(-\tau)$, and leads to
\be \rlo(t,\tw) = -e^{-\tau}\cg(\tw,\tw) + 1 - e^{-\tau} -
\tau\,e^{-\tau} \nn\ee
and hence\eq{eq:rlno_crit}.

The final case of interest is the dynamics at criticality but with
$m(0)\neq 0$. To obtain the leading contribution at long times to the
correction $\clo(t,\tw)$, consider the integral term
in\eq{eq:clo_explicit}. The non-local correlations are
$\cnl(t,\tw)=(\tw/2)(\tw/t)^{3/2}-\exp(-\tau)$.  The second term makes
a contribution which is at most $O(1)$, and exponentially suppressed
for $\tau\gg 1$. The first term, on the other hand, contributes

\be \int_{\tw}^t dt'\,e^{-(t-t')}\cg(t',\tw) \approx
\cg(t,\tw)\label{eq:clo_leading} \ee
where we have used that for $\tw\gg 1$ the factor $\cg(t',\tw)$ in the
integrand varies negligibly in the region $t-t'=O(1)$ that contributes
significantly, and the exponential integrates to unity for $\tau\gg
1$. Taking into account that the first term in\eq{eq:clo_explicit} is
also exponentially suppressed, equation\eq{eq:clo_leading} gives the
leading contribution to $\clo(t,\tw)$ for $\tau\gg 1$ and $\tw\gg
1$. For the response, very similar arguments show that the leading
contribution to the integral term in\eq{eq:rlo_explicit} is simply
$\rg(t,\tw)$; for time differences $\tau \gg 1$ the remainder of the
integral and the first term in\eq{eq:rlo_explicit} can be
disregarded. This leads to the results\eq{eq:clno_crit_m}
and\eq{eq:rlno_crit_m} given in the main text.

\begin{center}
\textbf{ACKNOWLEDGMENTS}
\end{center}

A.\ G.\ wishes to thank the University of Pompeu Fabra for its
support. F.\ R.\ is supported by the Ministerio de Eduacaci\'on y
Ciencia in Spain (BFM2001-3525), STIPCO (HPRN-CT-2002-00319), and by
the Distinci\'o de la Generalitat de Catalunya. I.\ P.\ acknowledges
financial support from DGICYT of the Spanish Government and from the
Distinci\'o de la Generalitat de Catalunya. All authors have been
supported by the SPHINX ESF program.

\end{document}